\documentclass[twocolumn,showpacs,preprintnumbers,amsmath,amssymb,floatfix,superscriptaddress]{revtex4}

\usepackage{amsbsy}
\usepackage{amsfonts}
\usepackage{amssymb}
\usepackage{amsmath}
\usepackage{graphicx,color}
\usepackage{graphicx}
\usepackage{epsfig}
\usepackage{graphicx} 
\usepackage{dcolumn}  
\usepackage{bm}       
\usepackage{color}
\usepackage{epstopdf}

\newcommand{\be}{\begin{equation}}
\newcommand{\ee}{\end{equation}}

\begin{document}

\title{ Bell inequality for pairs of particle number superselection rule restricted states }

\author{Libby Heaney}\email{l.heaney1@physics.ox.ac.uk}
\affiliation{Department of Physics, University of Oxford, Clarendon Laboratory, Oxford, OX1 3PU, UK}
\affiliation{Centre for Quantum Technologies, National University of
  Singapore, Singapore}

\author{Seung-Woo Lee}
\affiliation{Department of Physics, University of Oxford, Clarendon Laboratory, Oxford, OX1 3PU, UK}
\affiliation{Center for Subwavelength Optics and Department of Physics and Astronomy, Seoul National University, Seoul, 151-742, Korea}

\author{Dieter Jaksch}
\affiliation{Department of Physics, University of Oxford, Clarendon Laboratory, Oxford, OX1 3PU, UK}
\affiliation{Centre for Quantum Technologies, National University of
  Singapore, Singapore}

\begin{abstract}
Proposals for Bell inequality tests on systems restricted by the particle number superselection rule often require operations that are  difficult to implement in practice.  In this paper, we derive a new Bell inequality, where measurements on pairs of states are used as a method to by-pass this superselection rule.  In particular, we focus on  mode entanglement of an arbitrary number of massive particles and show that our Bell inequality detects the entanglement in an identical pair of states when other inequalities fail.   However, as the number of particles in the system increases, the violation of our Bell inequality decreases due to the restriction in the measurement space caused by the superselection rule.  This Bell test can be implemented using techniques that are routinely used in current experiments.
\end{abstract}

\maketitle

\section{Introduction}

The rise of quantum information science over the last decade has made it popular to seek and understand quantum many-body systems that contain entanglement \cite{Amico:08}.  One group of these many-body systems are the ultra-cold atomic gases, such as Bose Einstein condensates \cite{Pitaevskii:03}.
However, unlike spin lattice systems where the particles are distinguishable through position, particles in ultra-cold gases are typically indistinguishable from one another.    Indistinguishability means that the first quantised many-body state of the particles should be (anti)-symmetrized, but then the corresponding state space no longer has the tensor product structure required to define entanglement  \cite{Zanardi:01}.

One can, however, recover a tensor product state space by viewing the ultra-cold gases in terms of second quantised modes \cite{Peres:95}.  It has been shown that entanglement naturally exists between spatial modes  in non-interacting Bose Einstein condensates \cite{Simon:02, Toth:03, Anders:06} and in other ultra-cold atomic gases provided the coherence length of the particles extends over the selected modes \cite{Goold:09}.   In order to perform a Bell test on such systems, the spatial modes, which behave in some sense like a pair of qudits, must be rotated away from the particle number basis.  However, since systems of massive particles are restricted by a superselection rule \cite{Wick:52, Giulini:96, Wiseman:03} that forbids rotations away from the subspace of fixed particle number, a Bell-like test of the mode entanglement of massive particles is not straightforward. 
On the other hand, spatial mode entanglement (and non-locality) of a single photon  has been extensively studied \cite{Tan:91, Hardy:94, Greenberger:95, vanEnk:06, Dunningham:07, Heaney:09} and the experimental verification of single photon entanglement has been obtained \cite{Hessmo:04} via the CHSH Bell test \cite{Clauser:69}.  
\begin{figure}[b] 
   \centering
   \includegraphics[width=3.in]{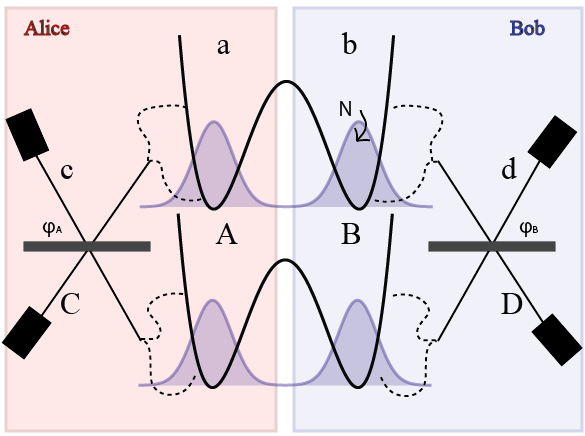} 
   \caption{Basic set up for the Bell test of mode entanglement using two copies, $\hat\rho_{aA}^{(N)}\otimes\hat\rho_{bB}^{(N')}$, of a bi-mode state  with $N$ and $N'$ massive particles respectively.  Modes $a$ and $A$ are given to Alice and $b$ and $B$ to Bob.  Each party makes a general measurement in the subspace of fixed particle number by sending their modes through a beamsplitter parameterised by the local angles $\varphi_A$ for Alice and $\varphi_B$ for Bob. }
   \label{fig:set-up}
\end{figure}  

Despite the superselection rule, a few schemes to test a Bell inequality with a single massive particle have recently been put forward \cite{Ashhab:07,Heaney:09a, Ashhab:09}.  In \cite{Heaney:09a} it was suggested that the spatial modes could be rotated away from the particle number basis by coupling to a coherent  particle reservoir.  However, to reliably confirm spatial mode entanglement of a single massive particle in an experiment,  one would have to ensure that no additional entanglement entered the system via the particle reservoir (see \cite{Williamson:09, Paterek:10} for further discussion of this point) and even if this were guaranteed,  such a coupling to the reservoir is likely to be difficult to implement in realistic conditions.  
Moreover, in both the photon experiment \cite{Hessmo:04} and in theoretical proposals \cite{Tan:91,Heaney:09a}, a local post-selection of the measurement outcomes was required to see a violation of the CHSH inequality, which meant that these tests only strictly probed for entanglement (i.e. the Bell inequality was used as an entanglement witness \cite{Guhne:09}), and not the more stringent property of non-locality.

In this paper, we derive a general Bell inequality to detect the mode entanglement of massive bosons, which overcomes the possible pitfalls pointed out above.  In particular, we use two copies of the system to make rotated measurements despite the superselection rule.  Using two copies not only eliminates the need to have careful couplings to a particle reservoir in order to overcome the superselection rule, but also ensures that no auxiliary entanglement from this reservoir will be responsible for the violation.  Note that Wiseman and Vacarro considered the entropic properties of pairs of superselection rule restricted states in \cite{Wiseman:03}.  Unlike previous tests of spatial mode entanglement that focus solely on single particles, our Bell inequality can be applied to systems containing any number of bosons, which can be both massless or massive.  At present there has been no experimental test for the existence of mode entanglement of massive particles -- even though this type of entanglement is predicted to be ubiquitous in ultra-cold atomic gases and that it has recently been  shown to be useful for quantum communication \cite{Heaney:09b}.  Because our test is relatively simple compared to previous proposals to detect mode entanglement of massive particles, its implementation could allow for the first direct confirmation of mode entanglement of massive particles.

We begin in the next section by explaining in more detail why two copies of a bi-mode quantum state allows one to make the measurements required to implement our Bell inequality despite the superselection rule.  Building on this, we derive in section (\ref{sec:derivation}) a general Bell inequality for two copies of a bi-mode state each with $N$ particles.  In section (\ref{sec:testing}) we test our Bell inequality with some examples of mode entangled states and in section (\ref{sec:discuss}) we discuss these results.  Finally, in section (\ref{sec:implementation}) we suggest how one could implement the test in practice, referring to current experimental techniques that could be put to use.

\section{The CHSH Bell inequality and the particle number superselection rule}\label{sec:ssr}

Bell inequalities allow to test the remarkable ability for entangled states to violate local realism.  Local realistic theories impose constraints on the correlations between measurement outcomes on two separated systems.
For systems occupying the state space, $\mathcal{H} = \mathcal{C}^{2}\otimes \mathcal{C}^{2}$,  the most commonly tested Bell inequality \cite{Bell:64} is the CHSH inequality \cite{Clauser:69}.  The CHSH inequality can be expressed in terms of joint expectation values of observables for two parties, $A$ and $B$, as
\begin{equation}
\label{eq:chsh}
\mathcal{B}_{CHSH}=|\langle \hat A_1 \hat B_1\rangle + \langle \hat A_1 \hat B_2\rangle  + \langle \hat A_2 \hat B_1\rangle - \langle \hat A_2 \hat B_2\rangle |\leq 2,
\end{equation}
where $\langle \hat A_i \hat B_j\rangle = \textrm{tr}[\hat A_i\hat  B_j\hat\rho_{AB}]$  and $\hat A_i$ and $\hat B_j$ each have two outcomes for $i,\,j = 1,2$.  When this inequality is violated there is entanglement between the two subsystems.


In systems of massive particles described by quantum mechanics, particle number is a superselected quantity so that the total particle number operator, $\hat N$, commutes with all other physical observables.  The system density operator, $\hat \rho$, therefore cannot contain any off-diagonal terms that connect states of different particle number.  The corresponding Hilbert space, $\mathcal{H}$, is decomposed as $\mathcal{H} = \oplus_{N=0}^\infty \mathcal{H}_N$, where $\mathcal{H}_N$ is the subspace containing $N$ particles.  For systems of two spatially separated modes, $A$ and $B$, the state space $\mathcal{H}_N$ has the substructure $\mathcal{H}_N=\oplus_{n=0}^N(\mathcal{H}_n^A\otimes\mathcal{H}_{N-n}^B)$, where $\mathcal{H}_n^{A(B)}$ denotes the Hilbert space of mode $A(B)$ with $n$ particles.

Now consider a single copy of a two mode system containing one massive particle.  A general state in the subspace, $\mathcal{H}_1$, spanned by $|01\rangle$ and $|10\rangle$ can be written as $\hat \rho_{AB} = p |\psi^+\rangle\langle\psi^+|_{AB} + (1-p) |\psi^-\rangle\langle\psi^-|_{AB}$, where $|\psi^{\pm}\rangle_{AB} = \frac{1}{\sqrt{2}}(|01\rangle\pm|10\rangle)$ and $|01\rangle= |0\rangle_A\otimes|1\rangle_B$ denotes no particles in mode $A$ and one particle in mode $B$.   The parameter, $p$, determines the entanglement of $\hat \rho_{AB}$, only when $p=\frac{1}{2}$ is the state separable, with $p=0$ or $1$ resulting in a maximally entangled state.  Because of the superselection rule, the only valid local measurement is particle number and, if we were to test Eq. (\ref{eq:chsh}), every joint correlation function, $\langle \hat A_i\hat B_j\rangle_{\rho_{AB}}$,   would be $\langle \hat A_i\hat B_j\rangle_{\rho_{AB}}= -1$ $\,\forall\, i, j $.  Since $\langle \hat A_i\hat B_j\rangle$ is independent of $p$, measurements in the particle number basis cannot distinguish separable states from entangled ones, and such measurements never allow for violation of the Bell inequality, even with a mode entangled state.  

Conversely, we take now two copies of the above state, $\hat \rho_{XY}$, shared between two parties, $\hat \sigma =\hat \rho_{AB} \otimes\hat \rho_{CD}$, where modes $A$ and $C$ are given to the first party and modes $B$ and $D$ to a second party. This composite state will allow to `see' the entanglement of $\hat \rho_{XY}$ despite the superselection rule.  
We consider now a toy example to highlight this point:   Each observer will make an (incomplete) local measurement on their two modes in the two-dimensional subspace of fixed particle number spanned by the normalised basis vectors, $\{ |\varphi+\rangle, |\varphi-\rangle\}$, where $|\varphi+\rangle = \alpha|10\rangle+\beta e^{i\varphi}|01\rangle$ and $|\varphi-\rangle = \beta e^{i\varphi} |10\rangle - \alpha |01\rangle$.  If the first party measures $\hat A(\varphi) = |\varphi+\rangle\langle\varphi+|- |\varphi-\rangle\langle\varphi - |$  and the second party $\hat B (\theta) = |\theta +\rangle\langle \theta +|-|\theta-\rangle\langle\theta -|$, the joint expectation values of Eq. (\ref{eq:chsh}) with the state, $\hat \sigma$, have the form, $\langle \hat A(\varphi) \hat B(\theta)\rangle_{ \hat \sigma} = 8 (p-\frac{1}{2})^2 \alpha^2\beta^2\cos(\varphi - \theta)$.  It is clear that the degree of correlation depends on the entanglement of the individual states, $\hat \rho_{XY}$, since the parameter, $p$, is present.  For instance, when the states are separable, i.e. $p=\frac{1}{2}$, the correlation function is $\langle \hat A(\varphi) \hat B(\theta)\rangle_{ \hat \sigma} = 0$, yet when $p = 0$ and $\alpha=\beta = \frac{1}{\sqrt 2}$, the correlation function is maximal, $\langle \hat A(\varphi) \hat B(\theta)\rangle_{ \hat \sigma} = \frac{1}{2}\cos(\varphi - \theta)$.  Moreover, the local parameters, $\varphi$ and $\theta$, can be altered by each party respectively to change between different measurement settings (1 and 2) required for a Bell test.  
We will now expand this basic example to formulate a general Bell inequality for pairs of $N$ particle states.

\section{Bell inequality for pairs of particle number superselection rule restricted states}\label{sec:derivation}

Consider two systems split into two spatially non-overlapping modes, the first system has $N$ massive particles and the second $N'$ massive particles.     We will denote the total state of the system, with $N+N'$ particles, by $\hat \sigma^{(N+N')} = \hat \rho_{ab}^{(N)}\otimes \hat \rho_{AB}^{(N')} $.  Party $A$ (or Alice) has access to modes, $a$ and $A$, and party $B$ (or Bob) to modes, $b$ and $B$, see Fig (\ref{fig:set-up}).

In our Bell test, Alice will make a joint measurement on her two modes in a subspace, $\mathcal{H}_M=\oplus_{n=0}^M(\mathcal{H}_n^{a}\otimes\mathcal{H}_{M-n}^{A})$, of fixed particle number, $M$, spanned by the basis states $\{ |M,0\rangle_{aA},|M-1,1\rangle_{aA},\dots,|0,M\rangle_{aA}\}$, although we do not know {\it a priori} what this number, $M$ ($0\leq M\leq N+N'$) will be.  
The  operator for this arbitrary high-dimensional measurement basis  is,  for Alice,
\begin{equation}
\label{eq:operator}
\hat A(\varphi_A^{(i)}) = \sum_{n_c+m_C=0}^{N+N'}\epsilon(n_c,m_C)|n_c,m_C\rangle\langle n_c,m_C|_{cC}.
\end{equation}
Likewise, Bob will  measure his two modes in the basis, $\hat B(\varphi_B^{(j)})$, and will receive $(N+N')-M$ particles. 
Here, $\epsilon (n_c, m_C)$ are weighting coefficients.  The local parameters, $\varphi_A^{(i)}$ and $\varphi_B^{(j)}$ where $i,\,j = 1,\,2$, denote the two measurement settings that Alice and Bob will use in the Bell test.  
Measuring each mode directly would allow for only local particle number measurements, but we would like to perform a general measurement within $\mathcal{H}_M$.  To do this each party passes their two spatial modes through a beamsplitter defined, for Alice, by the transformations 
\begin{equation}
\label{eq:bstransformations}
\hat c= \alpha\hat a+e^{-i\varphi_A}\beta \hat A\quad\quad \hat C = \beta \hat a - e^{-i\varphi_A}\alpha \hat A,
\end{equation}
where $\hat c$ and $\hat C$ are annihilation operators for the two output ports.  The operators $\hat a$ and $\hat A$ are annihilation operators for the two input modes of party $A$.  There are similar beamsplitter transformations for Bob, where we denote the output modes  as $\hat d$ and $\hat D$.     Each party measures the output modes in the particle number basis, the outcomes of which depend on the local parameter $\varphi_{A(B)}$.  Hence, the number of particles, $n_{c}$ and $m_{C}$, in the two output modes, $\hat c$ and $\hat C$, appear in the observable (\ref{eq:operator}) for Alice.
A measurement of $|n_c,m_C\rangle_{cC}$, corresponds to an effective measurement on the input modes, $a$ and $A$, of
\begin{eqnarray}
|n_c,m_C\rangle_{cC} &= &\frac{(\alpha \hat a^{\dag}+\beta e^{-i\varphi_A}\hat A^{\dag})^{n_c}}{\sqrt{n_c!}}\times\\
&&\quad\quad\quad\frac{(\beta \hat a^{\dag}-e^{-i\varphi_A}\alpha\hat A^{\dag})^{m_C}}{\sqrt{m_C!}}|0,0\rangle_{a,A},\nonumber
\end{eqnarray}
where $|0,0\rangle_{a,A}$ is the vacuum of modes $a$ and $A$ and we have used (\ref{eq:bstransformations}).  We will discuss in more detail a physical implementation of our test in section (\ref{sec:implementation}).

Since there are a total of $N+N'$ particles in the composite system, there are $(1/2(N+N')+1)(N+N'+1)$ number of different measurement
outcomes. For example, when $N=N'=1$ and for balanced beamsplitters, $\alpha=\beta=\frac{1}{\sqrt{2}}$, we have a total  of $6$ outcomes:
\begin{center}
\begin{tabular}{c|c|c}
 $|nm\rangle_{cC}$ & measurement on the modes $a$ and $A$ & $\epsilon(n_c,m_C)$ \\
\hline
  $|00\rangle$   &     $|00\rangle$ &1\\
   $|10\rangle$   &   $\frac{1}{\sqrt{2}}(|10\rangle+e^{-i\varphi_A}|01\rangle)$ &-1 \\
   $|01\rangle$   &   $\frac{1}{\sqrt{2}}(|10\rangle-e^{-i\varphi_A}|01\rangle)$ &1  \\
   $|20\rangle$   &   $\frac{1}{2}(|20\rangle+e^{-i\varphi_A}\sqrt{2}|11\rangle+e^{-i2\varphi_A}|02\rangle $ &-1 \\
   $|11\rangle$   &    $\frac{1}{\sqrt{2}}(|20\rangle-e^{-i2\varphi_A}|02\rangle)$ &1 \\
   $|02\rangle$   & $\frac{1}{2}(|20\rangle-e^{-i\varphi_A}\sqrt{2}|11\rangle+e^{-i2\varphi_A}|02\rangle$ &-1 \\

\end{tabular}\label{tab:1}
\end{center}
The weighting function, $\epsilon(n_c,m_C)$, is explicitly chosen as
\begin{equation}
\epsilon(n_c,m_C)=(-1)^{m_C+\frac{(m_C+n_c)(m_C+n_c+1)}{2}},
\end{equation}
which gives a sharp binning of results, as in \cite{Lee:09}.  There, sharp binning was shown not only to be optimal, but also to result in a tight Bell inequality.

We construct a Bell-type inequality from the local observables,
$\hat{A}(\varphi_A^{(i)})$ and $\hat{B}(\varphi_B^{(j)})$. Like the CHSH type
combination of Eq. (\ref{eq:chsh}), we formulate $
\hat{B}=\hat{E}(\varphi_A^{(1)},\varphi_B^{(1)})+\hat{E}(\varphi_A^{(1)},\varphi_B^{(2)})+
\hat{E}(\varphi_A^{(2)},\varphi_B^{(1)})-\hat{E}(\varphi_A^{(2)},\varphi_B^{(2)}),$
where the correlation operator is defined as
$\hat{E}(\varphi_A^{(i)},\varphi_B^{(j)})=\hat{A}(\varphi_A^{(i)})\otimes\hat{B}(\varphi_B^{(j)})$.
Since each local observable is bounded as $|\langle
\hat{A}(\varphi_A^{(i)}) \rangle |\leq1,\, |\langle \hat{B}(\varphi_B^{(j)}) \rangle |
\leq 1$, we obtain a Bell inequality from the expectation value of $\hat B$:
\begin{eqnarray}
\label{eq:BI}
|B_N|&=&|\mathrm{Tr}[\hat{\sigma}^{(N+N')}\hat{B}]|\\
&=&|E(\varphi_A^{(1)},\varphi_B^{(1)})+E(\varphi_A^{(1)},\varphi_B^{(2)})
+\nonumber\\
&&\quad\quad E(\varphi_A^{(2)},\varphi_B^{(1)})-E(\varphi_A^{(2)},\varphi_B^{(2)})|\leq 2,\nonumber
\end{eqnarray}
where the correlation function is 
\begin{eqnarray}
E(\varphi_A^{(i)},\varphi_B^{(j)}) =\sum_{\{n_c+m_C+n_d+m_D=N+N' \}}\epsilon(n_c,m_C)\times\nonumber\\
\epsilon(n_d,m_D)\,P^{(\varphi_A^{(i)},\varphi_B^{(j)})}(n_cm_C;n_dm_D).
\end{eqnarray}
Here $P^{(\varphi_A ^{(i)},\varphi_B ^{(j},)}(n_cm_C;n_dm_D)$
is the joint probability for the case that the outcome of
Alice and that of Bob is the trace of the
projection operator onto the states $|n_c,m_C\rangle_{cC}$ and
$|n_d,m_D\rangle_{dD}$ for the measurement setting $\varphi_A^{(i)}$ for Alice
and $\varphi_B^{(j)}$ for Bob. Therefore, if we can demonstrate the
violation of the given  Bell inequality (\ref{eq:BI}) for a quantum state,
$\hat{\sigma}^{(N+N')}$, then we can conclude that the state is entangled and, if the conditions for locality are met, also non-local.

\section{Applying the Bell inequality to pairs of mode entangled states}\label{sec:testing}

We will now evaluate the Bell inequality (\ref{eq:BI}) with different mode entangled states.    In the following section, we apply the Bell inequality to pairs of states of a zero temperature, non-interacting Bose Einstein condensate of fixed number of particles.  In section (\ref{sec:noonstates}), we apply the Bell inequality to states that are useful for precision measurement, such as the so called N00N states and the `spin' squeezed states.

\subsection{Non-interacting Bose-Einstein condensate}\label{sec:noninteract}

The zero temperature state of a non-interacting Bose-Einstein condensate of fixed particle number that is symmetrically distributed between two modes  is \cite{Simon:02}, 
\begin{equation}
\label{eq:psiN}
|\psi_N\rangle=\frac{1}{{\sqrt{2}}^N}\sum_{n=0}^N\frac{\sqrt{N!}}{\sqrt{n!(N-n)!}}|n,N-n\rangle.
\end{equation}\
Here we apply our Bell inequality (\ref{eq:BI}) to the pairs of states, $|\psi_N\rangle^{\otimes 2}$.
%

\begin{widetext}

\begin{figure}[htbp] 
   \centering
   \includegraphics[width=6in]{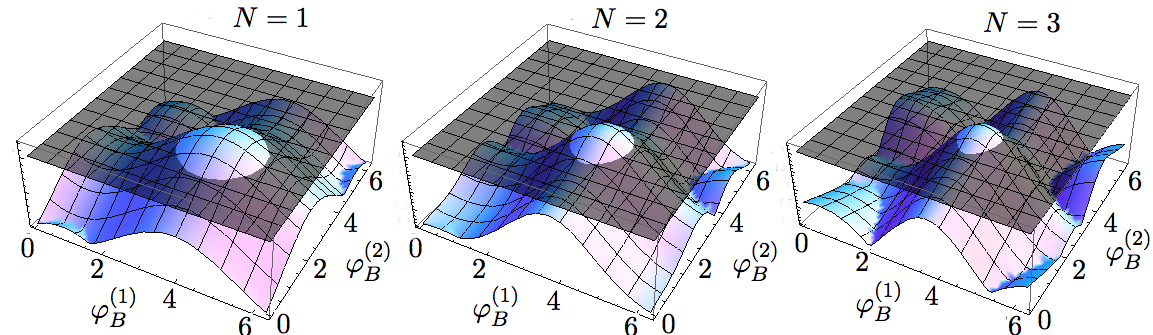}
   \caption{Three plots of the Bell terms, $B_{N=1}$, $B_{N=2}$ and $B_{N=3}$, with the state (\ref{eq:psiN}) from left to right.   
   Bell quantity $B_{N=1}$ has a maximum violation of $B_{N=1}=2.41$ at measurement settings: $\phi_A^{(1)}=0,\,\phi_A^{(2)} = 1.57,\,\phi_B^{(1)}=3.93$ and $\phi_B^{(2)} = 2.36$.  Bell quantity $B_{N=2}$ has a maximum violation of $B_{N=2}=2.36$ at measurement settings:  $\phi_A^{(1)}=0,\,\phi_A^{(2)} = 1.07,\,\phi_B^{(1)}=3.68$ and $\phi_B^{(2)} = 2.60$.   The final plot (on the right) is the Bell quantity $B_{N=3}$.  Here the maximum violation is $B=2.24$ at the angles of $\phi_A^{(1)}=0,\,\phi_A^{(2)} = 1.00,\,\phi_B^{(1)}=3.64$ and $\phi_B^{(2)} = 2.68$
   }
   \label{fig:N1}
\end{figure}

\end{widetext}

Let us first consider  the case of just a single particle ($N=1$), so that the composite state is $|\psi_1\rangle^{\otimes 2} = \left(\frac{1}{\sqrt{2}}(|10\rangle+|01\rangle)\right)^{\otimes 2}$.   The correlation function is $E_{N=1}(\varphi_A^{(i)},\varphi_B^{(j)}) =\sin^2((\phi_A^{(i)}-\phi_B^{(j)})/2)$, with the corresponding Bell term, $B_{N=1}$, constructed via Eq. (\ref{eq:BI}).
When $B_{N=1}>2$, the state distributed between Alice and Bob is non-local.  The left hand plot in figure (\ref{fig:N1}) shows on a violation of $B_{N=1}$ for a range of measurement settings.  We will discuss the results in more detail in the following section.

 We next consider $N=2$ particles in each system, so that the composite state is $|\psi_2\rangle^{\otimes 2}$.  Here the correlation function is $E_{N=2}(\varphi_A^{(i)},\varphi_B^{(j)}) =\sin^4((\phi_A^{(i)}-\phi_B^{(j)})/2)$, from which the Bell term, $B_{N=2}$, can be constructed.  
The centre plot in figure (\ref{fig:N1}) shows the Bell term, $B_{N=2}$.  While there is a violation of the Bell inequality, compared to the $N=1$ plot the maximum violation is smaller  and the range of measurement settings that give a violation has also reduced.  
For $N>2$ particles, the correlation functions become more complicated. For simplicity, we show only the plot of the Bell term, $B_{N=3}$, for the composite state, $|\psi_3\rangle^{\otimes 2}$, which is on the right hand side of figure (\ref{fig:N1}).
Again the magnitude of the violation and range of measurement settings that give a violation are smaller than in the cases of $|\psi_1\rangle^{\otimes 2}$ and $|\psi_2\rangle^{\otimes 2}$.  Note that for all the Bell inequalities in this section the value of the Bell term depends on the relative measurement settings between the parties, $\varphi_A^{(i)}-\varphi_B^{(j)}$ and not on the absolute phases of the two states.

We have also checked our Bell inequality with pairs of states, $|\psi_N\rangle\otimes|\psi_{N'}\rangle$, where $N\neq N'$ for balanced beamsplitters ($\alpha=\beta=\frac{1}{\sqrt{2}}$) on both sides and found the correlation functions are $E(\varphi_A^{(i)},\varphi_B^{(j)})=0$ for $(N,N') = (1,2),\,(1,3),\dots,(1,6),\,(2,3),\dots,(2,6),\,(3,4)$ irrespective of measurement settings.  For unbalanced beamsplitters ($\alpha\neq\beta$) on both sides, in general $E(\varphi_A^{(i)},\varphi_B^{(j)})\neq0$, but we have still found no violation of (\ref{eq:BI}) over all parameters for $(N,N') = (1,2)$.  Note that  $E(\varphi_A^{(i)},\varphi_B^{(j)})$ for $(N,N')$ is equal to that for $(N',N)$.

\subsection{Highly entangled states}\label{sec:noonstates}

We now apply the Bell inequality to states which contain more entanglement compared to the state of a non-interacting Bose Einstein condensate.  

\subsubsection{N00N states}
\begin{figure}[htbp] 
   \centering
   \includegraphics[width=3.3in]{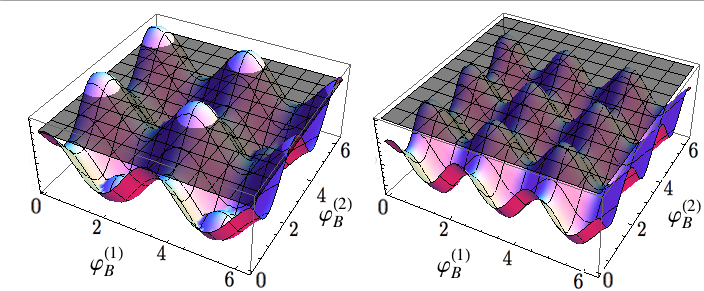}
   \caption{The left hand plot shows the identical Bell terms, $B_{N=2,m=0}$ and $B_{N=4,m=1}$, with the state (\ref{Eq:NmmN}).  The maximum violation here is $B_{N=2,m=0}=B_{N=3,m_1}=2.41$ at one of the four measurement settings: $\phi_A^{(1)}=-0.13,\,\phi_A^{(2)} =0.65,\,\phi_B^{(1)}=0.26$ and $\phi_B^{(2)} =-0.52$.  The right hand plot shows the Bell term, $B_{N=3,m=0}$, whose maximum value is $B_{N=3,m=0}=1.71$, hence there is no violation for all measurement settings.
   }
   \label{fig:N2m0}
\end{figure}

The so-called N00N states and their generalisations have been used, among other things, to gain drastic improvements in  precision measurements \cite{Giovannetti:04, Dorner:09, Jones:09} over the standard quantum limit.  Here we check our Bell inequality (\ref{eq:BI}) with pairs of the following states
\begin{equation}
\label{Eq:NmmN}
|N,m\rangle =\frac{1}{ \sqrt{2}}(|N-m,m\rangle+|m,N-m\rangle).
\end{equation}
Note that these states were also called MssM states in \cite{Jones:09} standing for ``{\it many-some + some-many}''.  

We first consider the composite state, $|N=2,\,m=0\rangle^{\otimes 2}=(1/\sqrt{2}(|20\rangle+|02\rangle))^{\otimes 2}$,    and compute the correlation function to be $E_{N=2,m=0}(\varphi_A^{(i)},\varphi_B^{(j)}) = \cos^2(\varphi_A^{(i)}-\varphi_B^{(j)})$.  The individual state, $|N=2,m=0\rangle$, was created in \cite{Folling:07} via second order tunneling. The corresponding Bell term, $B_{N=2, m=0}$, is plotted in figure (\ref{fig:N2m0}).  As with the case of $|\psi_1\rangle^{\otimes 2}$, the maximum violation here is $B_{N=2, m=0} = 2.41$; however, the number of regions of violation has increased from one to four, which corresponds to the enhanced phase sensitivity that such a state would bring in a precision measurement.   Moreover, the Bell term $B_{N=2, m=0}$ is identical to $B_{N=4, m=1}$ for the state $|N=4,\,m=1\rangle^{\otimes 2}=(1/\sqrt{2}(|31\rangle+|13\rangle))^{\otimes 2}$.
Conversely, we have checked our Bell inequality for the states, $|N=3,\,m=0\rangle^{\otimes 2}=(1/\sqrt{2}(|30\rangle+|03\rangle))^{\otimes 2}$, $|N=3,\,m=1\rangle^{\otimes 2}=(1/\sqrt{2}(|21\rangle+|12\rangle))^{\otimes 2}$ and $|N=4,\,m=0\rangle^{\otimes 2}=(1/\sqrt{2}(|40\rangle+|04\rangle))^{\otimes 2}$ and we have found that  there is no violation for all measurement settings.

\begin{figure}[htbp] 
   \centering
   \includegraphics[width=3in]{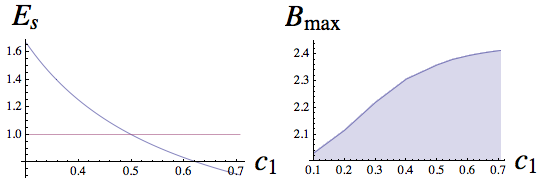} 
   \caption{The squeezing parameter, $E_S$, (left hand side) and the maximum violation of the Bell inequality (\ref{eq:BI}) (right hand side) for $c\leq1/\sqrt{2}$.  The state is spin squeezed when $E_S<1$.  Our Bell inequality detects weakly entangled states ($0\leq c< 1/2$) where the squeezing inequality (\ref{eq:squeezingineq}) does not.  }
   \label{fig:squeezevsbellmax}
\end{figure}

\subsubsection{Squeezed states}

\begin{widetext}

\begin{figure}[htbp] 
   {\centering
   \includegraphics[width=6 in]{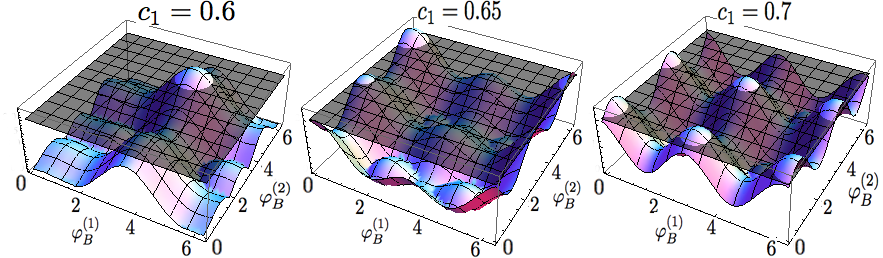} 
   \caption{The Bell terms (\ref{eq:BI}) for the state (\ref{Eq:squeezed}) for $c=0.6,\,c=0.65$ and $c=0.7$ as a function of two of the measurement settings (the other two are fixed to maximise the Bell term).  The maximum violation increases with increasing squeezing, namely $B_{N=2}(c=0.6)= 2.394,\,B_{N=2}(c=0.65)=2.405$ and $B_{N=2}(c=0.7)=2.413$.  The number of regions of violation grows with increasing squeezing, which is an indicator that the standard quantum limit in a precision measurement would be overcome.     }
   \label{fig:squeezegraph}}
\end{figure}

\end{widetext}

Spin squeezing \cite{Kitagawa:93,Sorensen:01} is a mechanism that generates states that surpass the standard quantum limit in precision measurements.   Here the $N$ particles, that are distributed between the two spatial modes, are described by a fictitious $J=N/2$ spin
 \cite{Bodet:10}.  The two modes would represent  the two states required to perform interferometry if that were our objective.  

Spin squeezing is achieved when the fluctuations in one angular momentum direction are reduced, while the coherence is preserved in at least one  of the other two directions.  We take here $\hat S_z = (1/2)(\hat a^{\dag}\hat a -\hat b^{\dag} \hat b),\,  \hat S_y = (i/2)(\hat a^{\dag}\hat b -\hat b^{\dag} \hat a)$ and $\hat S_x = (1/2)(\hat a^{\dag}\hat b+\hat b^{\dag}\hat a)$, where $\hat a$ and $\hat b$ are the annihilation operators for modes, $a$ and $b$.  The squeezing between the two modes is given by
\begin{equation}
\label{eq:squeezingineq}
E_S^2 = \frac{N(\Delta \hat S_z)^2}{\langle \hat S_x\rangle^2 +\langle \hat S_y\rangle^2},
\end{equation}
with $E_S<1$ corresponding to a spin squeezed state.

While in experiments \cite{Esteve:08} spin squeezing is generally achieved via a non-linear interaction with on order of $10^3$ particles, here we consider a toy example, which will allow comparisons to the other results in this section.  
We apply our Bell inequality to the composite state, $|\psi_2(c)\rangle^{\otimes 2}$, where
\begin{equation}
\label{Eq:squeezed}
|\psi_2(c)\rangle= c|20\rangle+\sqrt{1-2c^2}|11\rangle+c|02\rangle.
\end{equation}
The parameter, $c$, controls the amount of squeezing between the two modes; when $c=1/2$ the state is that of a non-interacting Bose-Einstein condensate, $|\psi_2\rangle$, and when $c=1/\sqrt{2}$ the N00N state, $|N=2,m=0\rangle$, from earlier in this section is reached.   The squeezing parameter, $E_S$, is plotted on the left hand side of figure (\ref{fig:squeezevsbellmax}) for different values of $c$.

Figure (\ref{fig:squeezegraph}) shows the Bell term (\ref{eq:BI}) for three different values of $c$ of increasing squeezing, namely $c=0.6, c=0.65$ and $c=0.7$.  For $c=0.6$, the Bell term still behaves in a similar manner to the $N=2$ non-interacting case from figure  (\ref{fig:N1}); there is one region of violation, but the squeezing has increased the maximum violation from $B_{N=2} = 2.36$ to $B_{N=2}(c=0.6)=2.394$.  The landscape of the Bell term, however, changes considerably as the squeezing gets larger still.  For $c=0.65$, there are two clear regions of violation indicating the amplitude of the $|11\rangle$ subspace is decreasing and the phase-enhancing $\{|20\rangle,\,|02\rangle\}$ subspace is playing a more significant role.  The maximum violation for $c=0.65$ is $B_{N=2}(c=0.65)=2.405$.  The final plot on the right-hand side of figure (\ref{fig:squeezevsbellmax}) shows the Bell term for a state which  predominantly consists of $|20\rangle+|02\rangle$, with a very small amount of $|11\rangle$ included.  Here we see four clear domains of violation and a maximum violation that is identical to the state $|N=2,\,m=0\rangle$.

\subsection{Weakly entangled states}

We can also apply our Bell inequality to weakly entangled states by taking the state (\ref{Eq:squeezed}) and allowing the parameter, $c$, to be set below the value for the non-interacting Bose-Einstein condensate case, i.e. $c$ will  be less than $1/2$.  By applying the standard von Neumann entropy to (\ref{Eq:squeezed}) one can detemine that  the state, $|\psi_2(c)\rangle$, becomes separable only when $c=0$.  Here the state is $|11\rangle$, which would correspond to the Mott regime of the Bose-Hubbard model.  

\begin{widetext}

\begin{figure}[h] 
   \centering
   \includegraphics[width=6 in]{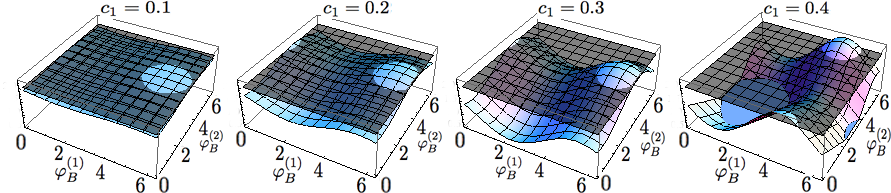} 
   \caption{(Color online) The Bell terms (\ref{eq:BI}) for the state (\ref{Eq:squeezed}) for $c=0.1,\,c=0.2,\,c=0.3$ and $c=0.4$ as a function of two of the measurement settings.  The entanglement of the state increases from left to right as does the range and the size of the violation of the Bell inequality.  Specifically the maximum violations for the four plots are $B(c=0.1) = 2.032,\, B(c=0.2) = 2.116,\, B(c=0.3) = 2.220$ and $B(c=0.4) = 2.307$.  }
   \label{fig:weakly}
\end{figure}

\end{widetext}

Figure (\ref{fig:weakly}) shows the Bell Term (\ref{eq:BI}) for the state $|\psi_2(c)\rangle^{\otimes 2}$ for values of $c$ where the entanglement is less than in the non-interacting BEC case.  For all values of $c$ there is just one region of violation.   As the entanglement in the state decreases, so does the maximum violation as can be seen more clearly in figure (\ref{fig:squeezevsbellmax}).  The Bell terms also become increasingly flattened with decreasing entanglement.   When $c=0$ the Bell term is flat and there is no violation of the Bell inequality.

\section{Discussion}\label{sec:discuss}
 
In the previous section, we applied the Bell inequality  (\ref{eq:BI}) to various mode entangled states of fixed particle number.  For the case of the non-interacting Bose Einstein condensate in section (\ref{sec:noninteract}), there was a decrease in both the size of the violation and in the range of measurement parameters over which a violation occurred with an increasing number of particles.  Restrictions on the measurement space due to the particle number superselection rule prevent the maximum violation ($2\sqrt{2}$) of (\ref{eq:BI})  occurring for any particle number, $N$, over any set of parameters, $\{\varphi_A^{(1)},\varphi_A^{(2)},\varphi_B^{(1)},\varphi_B^{(2)}\}$. We can explain this in the following way.

  Let us take the state, $|\psi_1\rangle^{\otimes 2}=(\frac{1}{\sqrt{2}}(|10\rangle+|01\rangle))^{\otimes 2}$. If on each run of our Bell test one particle were always guaranteed to end up with each party, then the maximum violation would be obtainable with the correct choice of measurement parameters.  One can see this by applying the Bell inequality (\ref{eq:BI}) to the renormalised state found by projecting $|\psi_1\rangle^{\otimes 2}$  onto the subspace spanned by $\{|1010\rangle_{aAbB},\,|1001\rangle_{aAbB},\,|0110\rangle_{aAbB},\,|0101\rangle_{aAbB}\}$.  
However, in practice there is always a finite probability for one party to detect both particles on their side of the system, which implements a local particle number measurement on each of the four modes $a,\,A,\,b$ and $B$ and will not distinguish between the quantum and classical correlations (see section (\ref{sec:ssr}) for a brief discussion of this point).    It is this mixing of outcomes from the two different measurement spaces that stops the Bell terms, $B_N$, from reaching their maximum value.

On the other hand, we know from Gisin and Peres \cite{Gisin:92} that for a spin singlet state of any size, $s$, one can always find measurement settings that give the maximum violation of $2\sqrt{2}$ to an  inequality identical to (\ref{eq:BI}).
The measurement operators used in \cite{Gisin:92}, that gave rise to the maximum violation, are block diagonal; with each  block consisting of a $2\times2$ rotation matrix, $\hat R_y (\alpha) = \hat \sigma_z \cos\alpha+\hat \sigma_x\sin\alpha$, where $\hat\sigma_z$ and $\hat \sigma_x$ are the usual Pauli operators.   This contrasts with our measurement operators (\ref{eq:operator}) that are block diagonal in the $n\times n$ subspaces of constant $n$ particles.     Since, due to the superselection rule, we cannot rotate our measurement operators with transformations that mix the subspaces of different particle number, it is impossible to reach the measurement space used in  \cite{Gisin:92} and hence it is also impossible to obtain the maximum violation of the Bell inequality.

Indeed, for a class of quantum optical down-conversion Bell tests, Popescu {\it et al.} \cite{Popescu:97} analyzed the CHSH inequality when  the measurement space also included outcomes that were unfavourableand they found that maximum violation of the inequality was $1+\sqrt{2}\approx2.41$.  This is identical to  the maximum violation we obtained in figure (\ref{fig:N1}) for $N=1$ and for the $|N=2,m=0\rangle$ state.

The decrease in the the range of violating measurement parameters with an increasing number of particles is also due to the restrictions on the measurement space.  This can be compared to an early result by Mermin \cite{Mermin:80}, who considered a Bell inequality for pairs of spin $s$ particles.  There Mermin created a restriction on his measurement space by considering only Stern-Gerlach type devices whose operation depends solely on the orientation of the quantization axis and thus cannot make projections onto arbitrary states of the subsystems.  In agreement with the results in section (\ref{sec:noninteract}), Mermin found that as the size of the spin increased, the range of angles for which the contradiction arose decreased.  
Speaking somewhat loosely, this restriction is similar to the fact that here when one party receives $M$ particles, the remaining $2N-M$ particles are {\it always} detected by the other party irrespective of the measurement setting chosen by each party.  Certain combinations of measurement outcomes are just impossible.  

In fact, Wiseman and Vaccaro suggested in \cite{Wiseman:03} that to correctly determine the entanglement for superselection rule restricted states one should first project such states into the subspace of fixed particle number, calculate the von Neumann entropy for resulting renormalised states and then take their average.  If one does this for pairs of states, $|\psi_N\rangle^{\otimes 2}$, the amount of  entanglement peaks for $N=2$ and then goes to zero by $N=9$.  Since we are also measuring in the subspaces of fixed particle number, we should likewise not expect to detect any entanglement for higher numbers of particles.  In order to see a maximum violation, one would need to make arbitrary measurements on the modes by coupling, for instance, to a Bose-Einstein condensate reservoir as in \cite{Heaney:09a, Heaney:09b}.

Our Bell inequality only shows violations for the pairs of N00N states with $N=2,\, m=0$ and $N=4, \, m=1$.  This is due to the measurements in our inequality being linear in particle number and including no higher order correlation functions.  As the basic group \cite{Yang:62} that forms the correlations in the N00N states increases, i.e. for $N-2m>2$, one would need second order observables (and higher) to detect the correlations.  In contrast to the non-interacting case, the entanglement of pairs of the N00N states as measured by Wiseman and Vaccaro's projected von Neumann entropy \cite{Wiseman:03} remains constant for all $N$, so that in principle the nonlocality of these states should be detectable within a different scheme.

We also checked the CGLMP inequality \cite{Collins:02} with the  joint probabilities, $P^{(\varphi_A ^{(i)},\varphi_B ^{(j},)}(n_cm_C;n_dm_D)$, for the non-interacting Bose-Einstein condensate and found no violation.  This suggests that our Bell inequality is particularly suited for detecting entanglement of pairs of states restricted by superselection rules.  Moreover, our Bell inequality is able to detect the mode entanglement of even weakly entangled states of two modes, $A$ and $B$ i.e. in state $|\psi_N\rangle$, when the well known spin squeezing inequalities \cite{Sorensen:01, Esteve:08} do not.

\section{Implementation with massive particles}\label{sec:implementation}

Finally, we discuss how to test this Bell inequality in realistic conditions.  For massive particles one can create the mode entangled state, $|\psi_N\rangle$, by cooling $N$ bosons into the ground state of a double well potential  \cite{Menotti:01, Shin:05}.  We require an identical pair of such  systems for our Bell test. The double wells would be positioned so that together they form a square like shape (see figure (\ref{fig:set-up}) for a rough indication of the set-up).  The potential barriers  between each of the wells would initially be high, while maintaining the coherence of the particles.  

To implement the beamsplitting operation each party lowers the potential barrier between for their wells for a desired time depending on the beamsplitter coefficients, $\alpha$ and $\beta$, generating an exchange of particles between modes, $a$ and $A$ \cite{Sengstock:04}.  
Similar beamsplitter networks were used in \cite{Alves:04} to detect multipartite entanglement between bosonic particles (as opposed to between bosonic modes, as in this paper), and subsequent work \cite{Palmer:05} showed that high precision beamsplitters for ultracold bosonic atoms can be realized based on current experimental technology.
 In our test, the different measurement settings are controlled by changing the relative phase between the two modes, $\hat a$ and $\hat A$, for Alice and $\hat b$ and $\hat B$ for Bob, by each party locally changing the bias of one mode (potential well) relative to the other, by applying, for instance, a dispersive laser pulse for a desired time \cite{Sengstock:04}.  
The number of particles in each of the wells is then counted.  While it is at the moment difficult to resolve different numbers of massive particles, steps in this direction have been made \cite{Schlosser:01}.  Each party would randomly choose a different measurement setting on each run of the test and the resulting value of the corresponding Bell term would be generated statistically over many runs.  Note that for massive particles it is unlikely that the measurements here would be performed at a speed faster than any communication between the modes, so that locality would not be assured, as is actually the case for all recent Bell tests with massive particles (note that a proposal was recently put forward for a loophole free Bell test using massive particles via entanglement swapping \cite{Rosenfeld:09}).


We note that one could also test this Bell inequality with mode entangled states of photons, which would allow the locality loophole to be closed.  To generate the entangled state, $|\psi_N\rangle$, one would send two lots of $N$ photons through two 50:50 beamsplitters, one output of each would be sent to Alice and the other output of each to Bob's side of the experiment.  These outputs would in turn then be passed through another beamsplitter, one for Alice and one for Bob, each with the desired reflectivity.  To switch between the measurement settings each party would pass one of their modes through a different phase-plate prior to the final beamsplitter.  Photon number would then be measured by each party in the output ports of the second set of beamsplitters and the magnitude of the Bell term generated over many runs of the test.  

\section{Conclusions}

In this paper, we have addressed the problem of performing a Bell inequality on quantum states that are restricted by the particle number superselection rule.  In particular, we have focused on states with a fixed number of massive particles, so that the particle number superselection rule is in effect.  We derived a Bell inequality that allows to by-pass the superseletion rule -- in order to perform measurements other than local particle number measurements, two copies of the states are used.  We test the Bell inequality with different mode entangled states and find that for a non-intereacting Bose-Einstein condensate, while the violation is not maximal, we detect some entangled states that the CGLMP Bell inequality cannot.  Moreover, the Bell inequality presented here can detect not only spin squeezed mode entanglement, but also relatively weak mode entanglement.
Our Bell inequality can be implemented with current technology.  

\section{Acknowledgements}

LH would like to thank D. Cavalcanti and W. Son for interesting discussions about this work.  LH acknowledges the support of the EPSRC, UK.  SWL acknowledges support of the National Research Foundation of Korea (NRF) grants funded by the Korean government (MEST) (R11-2008-095-01000-0/3348-20100018) and TJ Park Postdoctoral fellowship.
DJ and LH also acknowledge the National Research Foundation and the Ministry of Education (Singapore) for funding.

\end{document}